\documentclass {jltp}

\usepackage{graphicx}

\begin{document}

\title{Consideration of ac Josephson Effect in Fractional Quantum Hall States}

\author{Shosuke Sasaki}

\address {Shizuoka Institute of Science and Technology, Fukuroi, Shizuoka, 437-8555, Japan}

\maketitle

\begin{abstract}
The ac Josephson effect phenomenon is studied in fractional quantum Hall states (FQHS). Usual quantum Hall devices have an electron layer of uniform thickness. We consider a new quantum Hall device. The device has a narrow part that is vertical to the direction of the current. The narrow part has a smaller electron layer thickness than in the other part. The ac Josephson effect might be observed when the magnetic field or the electric current is modulated by an oscillation with a constant frequency value  $f$. Steps of the voltage appear when an electric current exceed a threshold value. 
The step value $V$ is related to the transfer charge $Q$ as $ V=(2 \pi \hbar f)/Q$.
We examine how the value of the transfer charge $Q$ depends on the fractional filling factor $\nu$. The $\nu$ - dependence of $Q$ is affected by the wave function of the fractional quantum Hall state. We classify the value of $Q$ according to the wave function type. Accordingly, observation of the ac Josephson effect in FQHS clarifies the structure of the fractional quantum Hall states.
\end{abstract}

\section{Introduction}
The fractional quantum Hall effect (FQHE) have been observed at fractional filling factors $\nu$ (=2/3, 3/5, 2/5 $\cdots$) in high mobility semiconductor heterojunctions. The confinements of the Hall resistance are extremely precise as clarified in experimental data. The FQHE phenomena are caused by Coulomb interactions among many electrons. Many physicists had interests to examine the origin of FQHE. 

R. B. Laughlin proposed an explicit trial wave function (Laughlin wave function). He argued that the elementary excitations are quasiparticles with fractional electric charge \cite{Landau}, and then explained the fractional quantum Hall effect. F. D. M. Haldane \cite{Haldane} and B. I. Halperin \cite{Halperin} extended the scheme of Laughlin. The quasiparticles and quasiholes have fractional charge and obey fractional statistics \cite{Zhang}. 
    
A model of composite fermions has been introduced by J. K. Jain \cite{Jain} and has been developed by many physicists \cite{Kamilla}, \cite{Blok}. The composite fermion consists of an electron (or hole) which is bound to an even number of magnetic flux quanta. Accordingly the composite fermions have integer charge, just like electrons. 

Recently, the perturbation energies via Coulomb transitions between many electrons are calculated for the FQH state with the minimum classical Coulomb energy \cite{SasakiNova}, \cite{Sasaki}, \cite{SasakiSpectrum}. In the result, the electron pair has a binding energy for specific filling factors. The typical energy spectrum has energy gaps at the specific filling factors as shown in Fig.1. We call the theory {\it electron pair theory of FQHE} hereafter.

\begin{figure}
\begin{center}
\includegraphics[width=16pc]{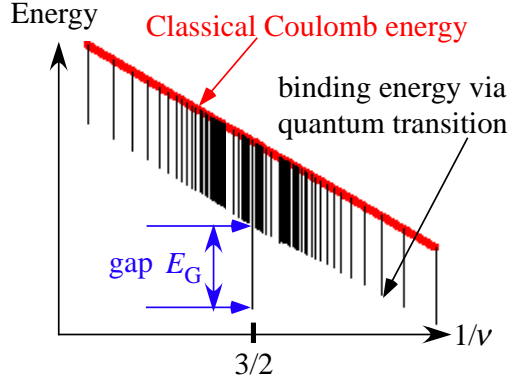}
\end{center}
\caption{\label{label}Energy spectrum of electron pair theory for FQH states near $\nu=2/3$. }
\end{figure}

This energy spectrum explains the precise confinement for plateau heights of Hall resistance at the specific filling factors \cite{SasakiSpectrum}. The mechanism is caused by the binding energies of electron pairs. 
These three types of theories have different wave functions. Accordingly the electric charge value of quasi-particle is different from each other in the three theories. At the filling factor $\nu$, the electric charge value $Q$ is

\begin{equation}
Q=\nu e 	 \hspace{2.4pc}		 \textrm {for Laughlin theory,}
\end{equation}
\begin{equation}		
\hspace{3pc}	Q=e 		 \hspace{2.4pc}		 \textrm {for composite fermion theory,}
\end{equation}
\begin{equation}	
\hspace{5pc}	Q=2e  	\hspace{2pc}		 \textrm {for electron pair theory of FQHE,}
\end{equation}
where the value $e$ expresses the elementary charge of electron. If we detect the charge value $Q$ through a Josephson junction in a quantum Hall device, then the mechanism of FQHE is more clarified. In this article, we propose an experiment detecting the transfer charge via a Josephson junction in a quantum Hall device.

\section{ac Josephson effect in quantum Hall device}

We consider the following experiment, namely measurement of ac Josephson effect in quantum Hall device. The device is schematically drawn in Fig.2. 
The central part of electron channel has a thickness less than in the other part as illustrated in the side view of Fig.2, where the 2D electron system is connected through this thin part from left to right. This type of Josephson junction is familiar in superconducting phenomena. When the current value exceeds the critical value of the central thin part, this thin part plays a role of Josephson junction. This type of Josephson junction is relatively-easy-to-make compared with original type of Josephson junction. 

\begin{figure}
\begin{center}
\includegraphics[width=20pc]{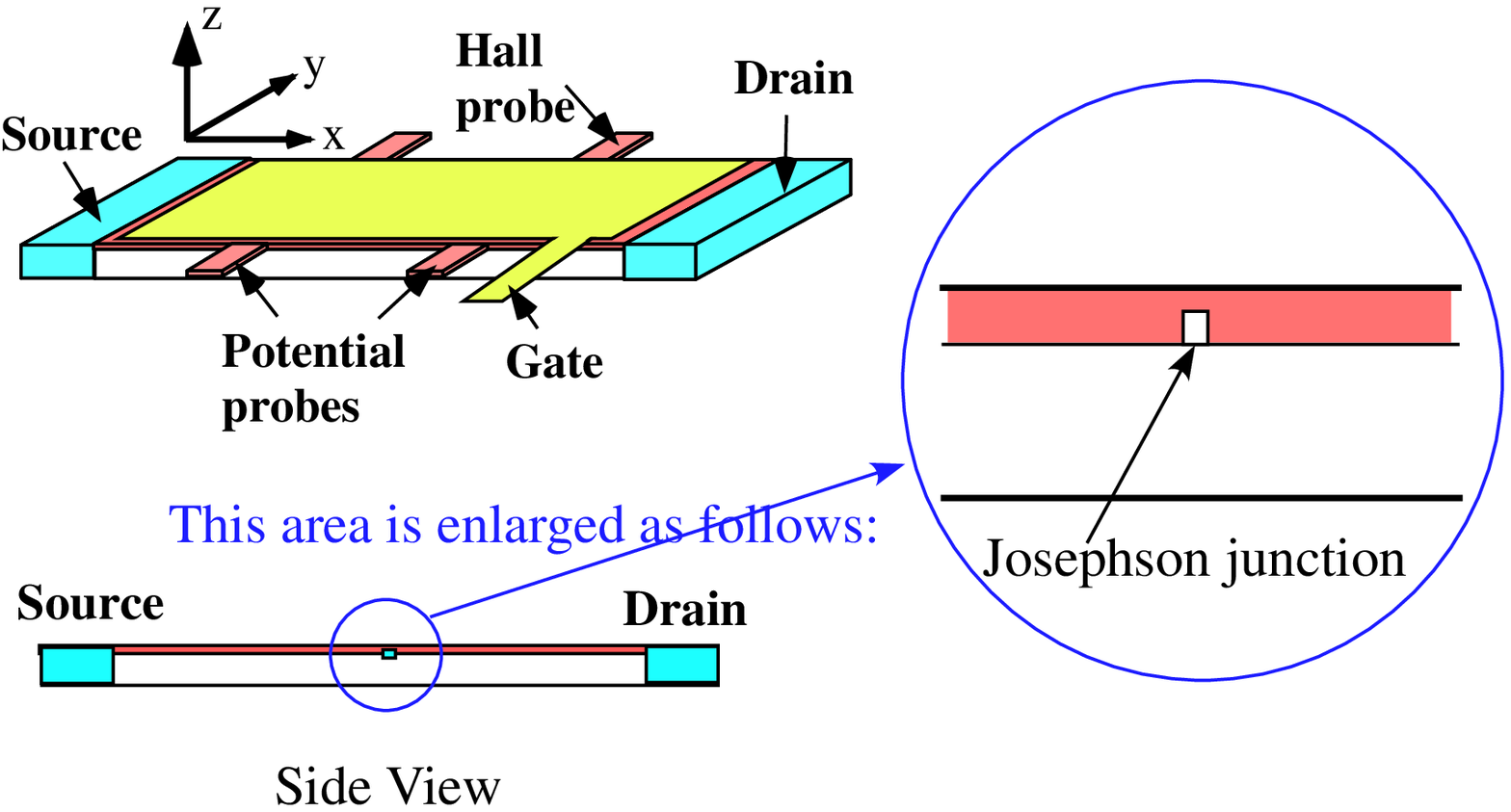}
\end{center}
\caption{\label{label}Josephson junction in quantum Hall device}
\begin{center}
\includegraphics[width=20pc]{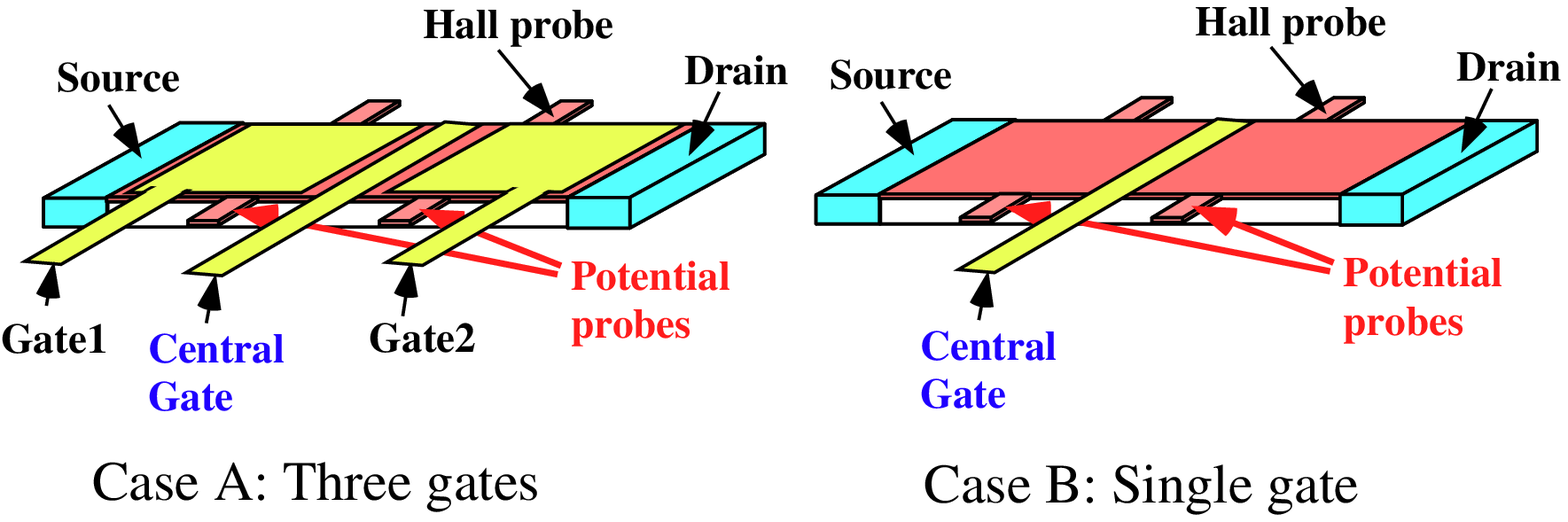}
\end{center}
\caption{\label{label}Different type of quantum Hall device with central gate}
\end{figure}

We can consider another type of device as illustrated in Fig.3. We can make small potential barrier, when the voltage of the central gate in Fig.3 is appropriately chosen. The small potential barrier plays a role of Josephson junction.

\section{Voltage Step depending on Quasi Charge}
The magnetic field is applied in the direction vertical to the 2D electron system. The strength of the magnetic field is adjusted to make an FQH state. Next, an oscillating magnetic field with the frequency value $f$ is added to the device. (We can use an oscillating current modulation on constant current instead of an oscillating magnetic field.) Thereafter we detect the voltage between two potential probes, while the electric current value between the source and the drain is changed from small to large. The voltage versus electric current value might have many steps. The predicted behaviors are schematically drawn in Fig.4. 
The value $e V/(h f)$ of the first step is equal to

\begin{equation}
e V/(h f)= e / Q
\end{equation}
where $V$ is the voltage value between two potential probes, $h$ is the constant of Planck. 
The value is 1.5 for Laughlin theory, 1 for composite fermion theory and 0.5 for electron pair theory at the filling factor $\nu=2/3$. Consequently this detection of voltage versus current is an interesting experiment. 

\begin{figure}
\begin{center}
\includegraphics[width=25pc]{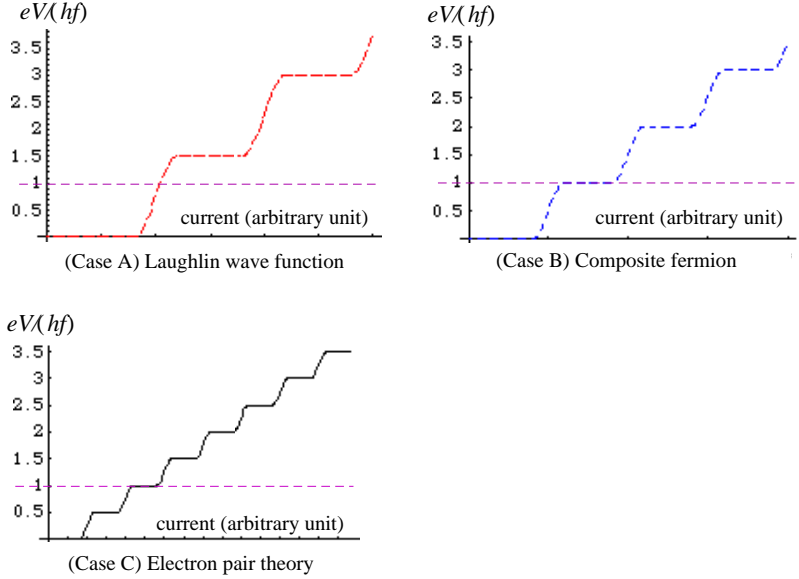}
\end{center}
\caption{\label{label}Diagonal voltage between potential probes in a quantum Hall device at the filling factor $\nu=2/3$ under the magnetic modulation with the oscillation frequency $f$.}
\end{figure}

\end{document}